\documentclass{PoS}

\title{Physics of top}

\ShortTitle{Physics of top}

\author{\speaker{C.-P. Yuan}
\\
Department of Physics $\&$ Astronomy\\
Michigan State University\\
East Lansing, MI 48824\\
USA\\

        E-mail: \email{yuan@pa.msu.edu}}

\abstract{ I will briefly review the physics of top quark at high
energy colliders. A new discovery of single-top event at the
Fermilab Tevatron is expected. At the CERN Large Hadron Collider,
detailed top quark properties can be measured and new physics
ideas in which top quark plays a special role can be tested. I
will also discuss a few phenomenological methods for analyzing
experimental data to study top quark interactions. }

\FullConference{International Workshop on Top Quark Physics\\

         January 12-15, 2006\\

         Coimbra, Portugal}

\begin{document}

\newcommand{\oalphas}{O(\alpha_{s})}
\newcommand{\met}{\not\!\! E_{T}}
\def\beq{\begin{equation}}
\def\eeq{\end{equation}}
\def\ba{\begin{array}}
\def\ea{\end{array}}
\def\bea{\begin{eqnarray}}
\def\eea{\end{eqnarray}}

\section{Introduction}

Prior to the discovery of top quark in 1995, a wide range of its
mass was predicted \cite{topmass}, which signals our ignorance
about the origin of mass. Hence, the breaking mechanisms of
electroweak symmetry (for generating $W$ and $Z$ boson masses) and
flavor symmetry (for generating a wide spectrum of fermion masses)
remain to be two of the major mysteries in the elementary particle
physics. Nevertheless, through the process of ``guessing'' what
the top quark mass is, we learned that only experimental data has
the final say about the mother Nature; the interaction between
experimentalists and theorists is essential for the advance of
science; and theorists should not give up any probable ideas. In
this talk, I will discuss a few theory ideas in which top quark
plays a special role such that studying the interaction of top
quark might help revealing the mechanism of electroweak symmetry
breaking (EWSB) and flavor symmetry breaking mechanisms. I will
also discuss a few phenomenological methods for analyzing
experimental data to study top quark interactions.

\section{Impact of $m_t$ measurement}

First, let us briefly review what we know so far about the top
quark. Its mass ($m_t$) has been measured with better accuracy
around 174\,GeV from studying the top quark pair events produced
via quark and gluon fusion processes at the Fermilab Tevatron
\cite{exptopmass}. It is quite challenging to measure $m_t$ from
the $bjj$ invariant mass to be better than a couple of GeV at the
Tevatron and the CERN Large Hadron Collider (LHC), because of the
uncertainty in jet energy resolution (related to under-lying
hadronic activities in the event) and the limited accuracy in
current theory calculations when the accuracy in the $m_t$
measurement ($\delta m_t$) is required to be less than the width
of top quark. Other means for measuring $m_t$ are: studying the
fraction of longitudinal polarization of the $W$ boson (or the
transverse momentum ($p_T$) distribution of the lepton) and the
invariant mass distribution of $b$ and $\ell$ in the decay of top
quark $t \rightarrow b W (\rightarrow \ell \nu)$ \cite{kane}. How well do we need
to know about $m_t$ in order to test the Standard Model (SM)
beyond the tree level? At the Tevatron Run-II, $\delta m_t \sim 2
- 3$\,GeV, and it is no longer a dominant error in precision test
until $\delta M_W$ is reduced to about 20\,MeV. At the LHC,
$\delta m_t \sim 1.5$\,GeV, which is about the same as the top
quark width ($\Gamma_t$), and the counterpart in the measurement
of $M_W$ needs to be about 10\,MeV. Again, we see that $\delta
m_t$ is not the dominant theoretical error in testing the SM from
comparing its predictions to precision data.

At the future International Linear Collider (ILC), $\delta m_t$
can be reduced to a couple of hundred MeV (of the order of
$\Lambda_{QCD}$) from studying the top quark pair production at
threshold (from the measurement of total cross section, peak of
transverse momentum distribution, and the forward-backward
asymmetry), and around 500\,MeV from directly reconstructing the
mass of top quark produced at continuum \cite{ILC}.

\section{Top quark decay}

Because its mass is at the weak scale, top quark will decay like a
bare quark (without first forming hadrons) so that we could study
its spin property in addition to measure its decay branching
ratios. In the SM, top decays into $bW$ mode almost 100\% of time.
With new physics coupling to top quark, many new decay channels
become possible, such as $t \rightarrow bH^+, {\tilde t} \chi_0, c \gamma,
c Z, c g, c h^0$ predicted in the Minimal Supersymmetric Standard
Model (MSSM). Thus, it is important to measure the decay branching
ratio (BR) of $t \rightarrow b W$. Unfortunately, in the SM, the CKM
matrix element $V_{tb}$ is so much larger than $V_{ts}$ and
$V_{td}$ that the ratio of BR($t \rightarrow b W$) to BR($t \rightarrow b q$)
cannot effectively measure the magnitude of the $W$-$t$-$b$
coupling. Moreover, the total decay width of the top quark cannot
be accurately measured from the $bjj$ invariant mass distribution
in top decays, for the experimental uncertainty due to jet energy
resolution is much larger than $\Gamma_t$ \cite{mrenna}.

It is however possible that new physics effect might not change
the value of BR($t \rightarrow b W$), for not having additional new light
fields with mass less than $m_t$, but could still modify
$\Gamma_t$ when the interaction of $t$-$b$-$W$ is strongly
modified. Hence, we need to directly measure the interaction
strength of $t$-$b$-$W$ coupling. This can be done by studying
single-top production rates initiated from weak charged current
processes. In particular, the SM tree level t-channel single-top
inclusive production rate ($\sigma_t$) is proportional to the
decay width of top quark. Hence, $\sigma_t$ can be used to
determine the CKM matrix element $V_{tb}$ in the SM, and to
determine the partial decay width $\Gamma(t \rightarrow b W)$
\cite{larios}. When combining the measurement of BR($t \rightarrow b W$)
from $t \bar t$ events and $\Gamma(t \rightarrow b W)$ from t-channel
singe-top event, one can determine the life-time of top quark from
their ratio.

\section{Single-top production}

At the Tevatron and the LHC, single-top events can be produced
from t-channel, s-channel and $Wt$ associate production processes
\cite{tait}. Their production rates can be largely modified by new
physics interactions, either with new heavy resonances, such as
$W',Z',H^\pm,{\pi}^\prime $, or with flavor changing neutral
currents (FCNC), such as $tcZ,tuZ,tcg,tc\gamma$, or with new
flavor changing charged currents (FCC), such as $tsW,tdW,tbH^+$.
It turns out that the s-channel mode is sensitive to charged
resonances. The t-channel mode is more sensitive to FCNC and new
interactions. The $t W$ mode is a more direct measure of top quark
coupling to $W$ and a down-type (down, strange, bottom) quark,
such as FCC couplings. From a theoretical point of view, they are
sensitive to different new Physics. From an experimental point of
view, they have different signatures and different systematics
\cite{tait}. Hence, they should be separately measured
experimentally. Furthermore, as a proton-antiproton ($p \bar p$)
collider, Tevatron offers a special chance to measure the amount
of (direct) CP-violation in top quark system by measuring the
asymmetry in the inclusive single-top versus single-antitop
production rates \cite{cpy}: \beq A_t^{CP}={ {\sigma(p {\bar p}
\rightarrow tX) - \sigma(p {\bar p} \rightarrow {\bar t} X) } \over {\sigma(p
{\bar p} \rightarrow tX) + \sigma(p {\bar p} \rightarrow {\bar t} X) }} \, . \eeq
This is because under the CP symmetry, the initial hadron state
($p \bar p$) is invariant. This is the unique opportunity at
Tevatron that the LHC cannot offer. To probe CP-violation in top
quark interaction at the LHC, one has to measure the CP-violating
(more generally, time-reversal violating) observables that make
use of the spin information of the produced (anti)top quarks, such
as measuring the expectation values of $<{\vec s}_t \cdot {\vec
p}_b \times {\vec p}_{\ell^+}>$ and $<{\vec s}_{\bar t} \cdot
{\vec p}_{\bar b} \times {\vec p}_{\ell^-}>$ from the decay of top
and anti-top, respectively, in the single-top and single-antitop
events \cite{cpy}. Needless to say that CP-violation can also be
tested at the LHC in the $t \bar t$ events by comparing the
production rates of $t_L {\bar t}_L$ and $t_R {\bar t}_R$ events,
for under the CP operation, a left-handed top ($t_L$) becomes a
right-handed anti-top (${\bar t}_R$). One way to measure this
asymmetry is to detect the asymmetry in the energy distributions
of $\ell^+$ versus $\ell^-$ in the inclusive $t \bar t$ events
\cite{schmidt}.

\section{Single-top production and decay at NLO QCD}

In order to reliably compare the theory prediction with
experimental data on the production rate of the single-top events
and the distributions of its final state particles, a
next-to-leading order (NLO) QCD calculation has been performed for
the s- and t-channel single-top processes \cite{nloqcd}. In
Ref.~\cite{nloqcd}, we separated the single-top processes into a
few smaller gauge invariant sets to organize our calculations,
which include corrections to the initial state, final state and
top decay in the s-channel process, and corrections to the light
quark line, heavy quark line and top decay in the t-channel
process. Keeping track on each individual contributions is useful
for comparing event generators with exact NLO predictions. One can
also study the effect from, for example, having correct
implementation of top quark spin correlation between its
production and decay at the full NLO in QCD.

Given the small single-top production rate at the Tevatron, as
predicted by the SM, it is important to study the acceptance of
the signal event after imposing the necessary kinematic cuts for
detecting them experimentally. We found that the signal
acceptances are sensitive to the kinematics cuts. A large
$R$-separation cut reduces acceptances significantly because
$\Delta R( \ell j)$ is typically less than 1. With tight kinematic
cuts, leading order (LO) and NLO acceptances are almost the same,
but, with lose cuts, they are quite different. Hence, in order to
maximize the signal acceptance, we must impose lose cuts and
consequently the NLO acceptance of the signal events cannot be
correctly modelled by a scaled-up (multiplied by the $K$-factor
extracted from inclusive rate calculations) tree level LO
calculation.

To identify the single-top signal events and to test the
polarization of top quark, by studying spin correlations among the
final state particles, we need to reconstruct the top quark in the
single-top events. To do so, we shall first identify the $b$-jet
and reconstruct the $W$ boson from top decay. In
Table~\ref{tab:efficiency-bnu}, we show the efficiency of finding
the correct $b$-jet ($\epsilon_{b}$) in two different algorithms:
best-jet algorithm and leading $b$-tagged jet algorithm. The
{}``best-jet'' is defined to be the $b$-tagged jet which gives an
invariant mass closest to the true top mass when it is combined
with the reconstructed $W$ boson after determining the
longitudinal momentum $p_{z}^{\nu}$ of the neutrino from $W$
decay. The leading $b$-tagged jet algorithm picks the leading
$b$-tagged jet as the correct $b$-jet to reconstruct the top quark
after combining with the reconstructed $W$ boson. As shown in
Refs.~\cite{Cao:2004ap,Cao:2005pq}, we find that the best-jet
algorithm shows a higher efficiency (about $80\%$) in picking up
the correct $b$-jet than the leading-jet algorithm (about $55\%$)
for the $s$-channel single-top events. On the other hand, for the
$t$-channel single-top events, the leading $b$-tagged jet
algorithm picks up the correct $b$-jet with a higher efficiency,
about $95\%$ for inclusive 2-jet events and $90\%$ for exclusive
3-jet events., cf. Table 1. The reason that the leading $b$-tagged
jet algorithm works well in the exclusive 3-jet $t$-channel
single-top events is due to the distinct kinematic difference
between $b$ and $\bar{b}$-jets. To reconstruct the top quark in
the signal events, we also need to reconstruct the $W$ boson. The
$W$ boson is reconstructed with the help of using its mass
constraint: $M_{W}^{2}=(p_{l}+p_{\nu})^{2}$. Which one of the
two-fold solutions in $p_{z}^{\nu}$ to be taken depends on the
$b$-jet algorithm we used. In the case of best-jet algorithm, we
find the one with the smaller magnitude from solving the $W$ mass
constraint give the best efficiency in $W$ boson reconstruction.
In the case of leading $b$-tagged jet algorithm, we use the top
quark mass constraint $M_{t}^{2}=(p_{b}+p_{l}+p_{\nu})^{2}$ to
pick up the best $p_{z}^{\nu}$ value. The efficiency for picking
up the correct $p_{z}^{\nu}$ value ($\epsilon_{\nu}$), at LO and
NLO, respectively, is presented in Table~\ref{tab:efficiency-bnu}.

\begin{table}
\begin{tabular}{c|c|c|c|c|c|c}
\hline & \multicolumn{3}{c|}{best-jet algorithm}&
\multicolumn{3}{c}{leading $b$-tagged jet
algorithm}\tabularnewline \hline & $s$-channel&
\multicolumn{2}{c|}{$t$-channel}& $s$-channel&
\multicolumn{2}{c}{$t$-channel}\tabularnewline \hline & inclusive
2-jet& inclusive 2-jet& exclusive 3-jet& inclusive 2-jet&
inclusive 2-jet& exclusive 3-jet\tabularnewline \hline
$\epsilon_{b}$& $80\%$& $80\%$& $72\%$& $55\%$& $95\%$&
$90\%$\tabularnewline \hline $\epsilon_{\nu}$&
\multicolumn{3}{c|}{$70\%$}&
\multicolumn{3}{c}{$84\%$}\tabularnewline \hline
\end{tabular}

\caption{Efficiencies of identifying correct $b$-jet
($\epsilon_{b}$) and picking up correct $p_{z}^{\nu}$
($\epsilon_{\nu}$) in both best-jet algorithm and leading-jet
algorithm. \label{tab:efficiency-bnu}}
\end{table}

After reconstructing the top quark, one can study the effect of
NLO QCD corrections to the measurement of top quark polarization.
We found that higher order QCD corrections blur the spin
correlation effect. Furthermore, the apparent advantage of some
polarization basis at the parton level is washed away after
modelling the reconstruction of the top quark as described above.

\subsection{s-channel single-top events and Higgs search}

\begin{figure}
\includegraphics[%
  scale=0.4]{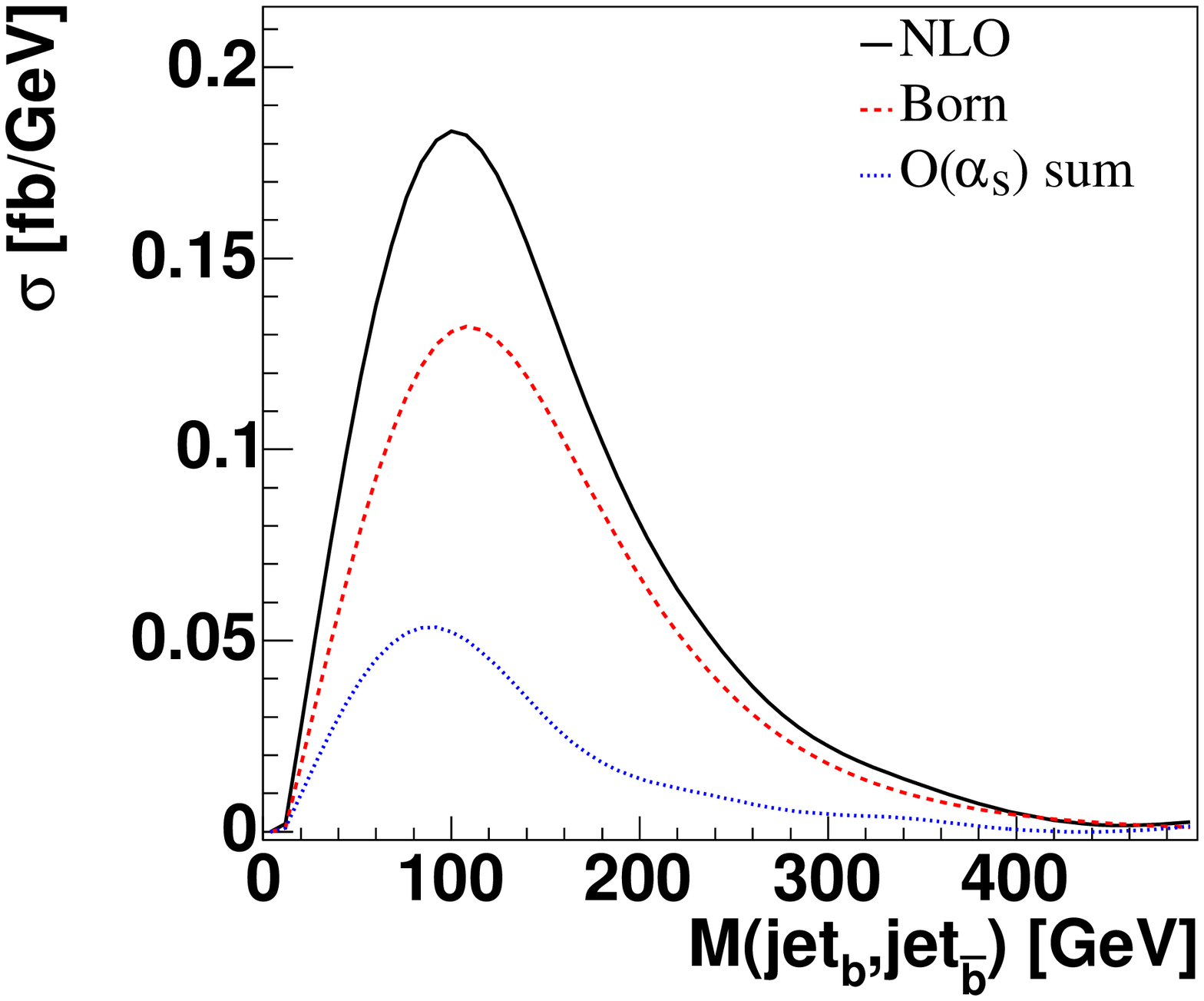}\hspace*{-4mm}\includegraphics[%
  scale=0.4]{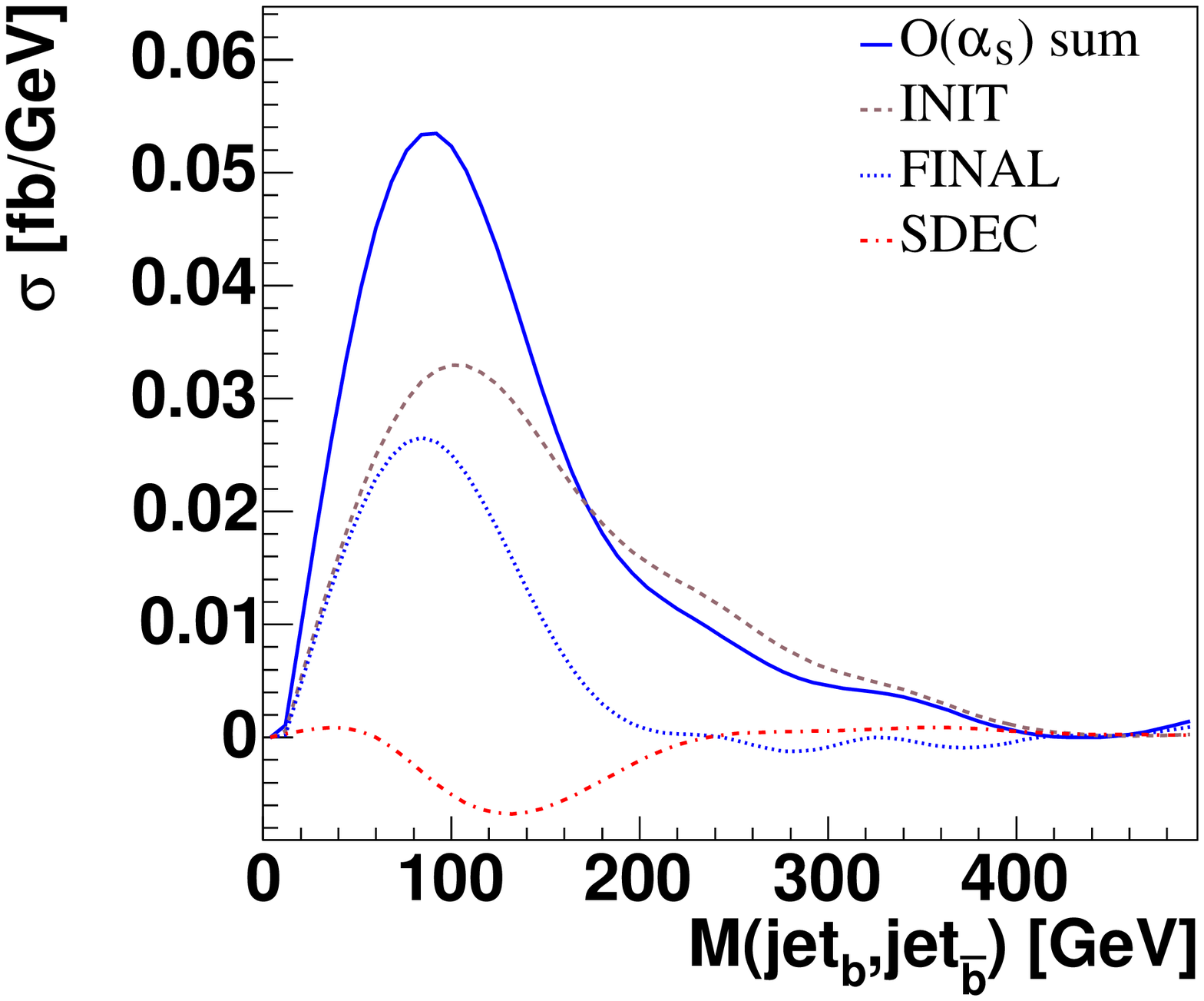}

\caption{Invariant mass of the ($b$-jet, $\bar{b}$-jet) system
after selection cuts, comparing Born-level to $\oalphas$
corrections. In the legend, INIT, FINAL and SDEC denotes the
contributions from initial state, final state and top quark decay
corrections, respectively.\label{fig:bJetbbarJetMass-schan}}
\end{figure}

 The $s$-channel single top quark process also
contributes as one of the major backgrounds to the SM Higgs
searching channel $q\bar{q}\rightarrow WH$ with $H\rightarrow
b\bar{b}$. In this case it is particularly important to understand
how the $\oalphas$ corrections change distributions around the
Higgs mass region. Because of the scalar property of the Higgs
boson, its decay products $b$ and $\bar{b}$ have symmetric
distributions. Fig.~\ref{fig:bJetbbarJetMass-schan} shows the
invariant mass distribution of the ($b$-jet, $\bar{b}$-jet)
system. For a Higgs signal, this invariant mass of the two
reconstructed $b$-tagged jets would correspond to a plot of the
reconstructed Higgs mass. Thus, understanding this invariant mass
distribution will be important to reach the highest sensitivity
for Higgs boson searches at the Tevatron. The figure shows that at
$\oalphas$, the invariant mass distribution not only peaks at
lower values than at Born level, it also drops off faster. This
change in shape is particularly relevant in the region focused on
by SM Higgs boson searches of $80\,{\rm GeV}\leq
m_{b\bar{b}}\leq140\,{\rm GeV}$ which is also at the {\it fb}
level. In particular, the NLO contribution from the decay of top
quark, while small in its overall rate, has a sizable effect in
this region of the invariant mass and will thus have to be
considered in order to make reliable background predictions for
the Higgs boson searches.

Other kinematic distributions are also changing in shape when
going from Born-level to $\oalphas$.
In Ref.~\cite{Cao:2004ap}, we also showed the distribution of
$\cos\theta$ for the two $b$-tagged jets, where $\theta$ is the
angle between the direction of a $b$-tagged jet and the direction
of the ($b$-jet, $\bar{b}$-jet) system, in the rest frame of the
($b$-jet, $\bar{b}$-jet) system. Experiments cannot distinguish
between the $b$- and the $\bar{b}$-jets, we therefore include both
the $b$-jet and the $\bar{b}$-jet in the graph. This distribution
is generally flat at Born-level, with a drop-off at high
$\cos\theta$ due to jet clustering effects, and a drop-off at
negative $\cos\theta$ due to kinematic selection cuts. The
$\oalphas$ corrections change this distribution significantly and
result in a more forward peak of the distribution, similar to what
is expected in Higgs boson production. In other words, a flatter
$\cos\theta$ distribution in the s-channel single-top events make
it more difficult to separate the $WH$ events from the s-channel
single-top events in experimental analysis.

\subsection{t-channel single-top events and Higgs search}

The unique signature of the $t$-channel single top process is the
spectator jet in the forward direction, which can be utilized to
suppress the copious SM backgrounds, such as $Wb\bar{b}$ and
$t\bar{t}$ events \cite{st1990}. Studying the kinematics of this
spectator jet is important to have a better prediction of the
acceptance of $t$-channel single top quark events. The impact of
the NLO QCD corrections on the kinematic properties of the
spectator jet has been reported in Ref.~\cite{Cao:2005pq} for the
Tevatron Run-II phenomenology. We found that the NLO QCD
corrections to the light-quark (LIGHT) and heavy-quark (HEAVY)
lines show almost opposite behavior in the rapidity distribution
of the spectator jet. LIGHT shifts the spectator jet to the
forward direction while HEAVY shifts it to the central region, and
the NLO corrections originated from the decay of top quark does
not modify the rapidity distribution of the spectator jet.

One of the most important tasks at the LHC is to find the Higgs
boson, denoted as $H$. It has been shown extensively in the
literature that the production mechanism of Higgs boson via weak
gauge boson fusion is an important channel for the search of Higgs
boson. Furthermore, to test whether it is a SM Higgs boson after
its discovery, one needs to determine the coupling of $H$-$V$-$V$,
where $V$ denotes either $W^{\pm}$ or $Z$, by measuring the
production rate of $q\bar{q}(VV)\rightarrow Hq^{\prime}\bar{q}^{\prime}$
via weak boson fusion processes. In order to suppress its large
background rates, one usual trick is to tag on the two
forward-jets resulted from emitting vector boson $V$ to produce
Higgs boson via $VV\rightarrow H$. Prior to the discovery of Higgs boson,
one can learn about the detection efficiency of the forward jet
from studying the $t$-channel single-top process. This is because
in the $t$-channel single-top process, the forward jet also
results from emitting a $W$ boson which interacts with the $b$
quark from the other hadron beam to produce the heavy top quark.
As pointed out in Ref.~\cite{st1990}, in the effective-$W$
approximation, a high-energy $t$-channel single top quark event is
dominated by a longitudinal $W$~boson and the $b$~quark fusion
diagram. It is the same effective longitudinal $W$~boson that
dominates the production of a heavy Higgs boson at high energy
colliders via the $W$-boson fusion process. For a heavy SM Higgs
boson, the longitudinal $W$~boson fusion process dominates the
Higgs boson production rate. Hence, it is important to study the
kinematics of the spectator jet in $t$-channel single top quark
events in order to have a better prediction for the kinematics of
Higgs boson events via the $WW$ fusion process at the LHC.

\section{General Analysis of $t$-$b$-$W$ couplings}

As discussed above, the $t$-$b$-$W$ couplings can affect the spin
correlations among the decay particles of (anti)top quarks
produced in the $t \bar t$ events and the production cross section
of the single-top events. Therefore, it is desirable to find a
systematic way to analyze both $t \bar t $ and single-top
experimental data to extract out the information on the
$t$-$b$-$W$ couplings. This was performed in Ref.~\cite{larios},
in which the most general formulation of the $t$-$b$-$W$ couplings
in any weak-scale effective theory was formulated. It also
summarized what we have learned from indirect measurements whose
conclusions unavoidably depend on the specific assumption made
about the underlying new physics models. The most interesting
question to ask is ``How to directly measure $t$-$b$-$W$ couplings
at the Tevatron and the LHC?'' as well as ``How to distinguish
models of EWSB from the results of these direct measurements?''.

In the language of electroweak chiral Lagrangian, there are in
general 8 different operators at the weak scale to describe the
$t$-$b$-$W$ couplings which do not require either $t$, $b$ or $W$
to be on their mass-shells. For on-shell $t$ and $b$ quarks, the
above 8 operators reduce to 6 independent ones. Since we will
consider only the case that the $W$ boson in the $t$-$b$-$W$
couplings couples to massless fermions (quarks or leptons), the
number of independent operators is reduced to 4. In the Unitary
gauge, it reads as \cite{kane} \bea \mathcal{L}_{\mathrm tbW} &=&
\frac{g}{\sqrt 2}\, W^-_\mu \, \bar b \, \gamma^\mu  \left( f_1^L
P_L + f_1^R P_R  \right)\, t
- \frac{g}{\sqrt 2 M_W} \,
\partial_\nu W^-_\mu \, \bar b \, \sigma^{\mu\nu}
\left( f_2^L P_L + f_2^R P_R \right) \, t \;\;\;+\; h.c.\, .
\label{tbwvertex} \eea
At tree level, the SM predicts $f^L_1 = V_{tb}\simeq 1$, and
$f^R_1 = f^L_2 = f^R_2 = 0$. Since these form factors can take
different values depending on the underlying new physics models
that lead to the above effective Lagrangian at the weak scale, we
proposed a general analysis to determine these four independent
$t$-$b$-$W$ couplings. The idea is to use four experimental
observables to determine the four independent form factors
$f_{1,2}^{L,R}$. They are the experimental measurements on the
degrees of longitudinal ($f_0$) and left-handed ($f_-$)
polarizations of the $W$ bosons from top decays measured in the $t
\bar t$ events and the s- ($\sigma_s$) and t-channel ($\sigma_t$)
single-top production rates.

In summary, the above four observables can be expressed in terms
of the effective $t$-$b$-$W$ couplings as:\footnote{There are
typos in Eqs. (5) and (6) of Ref.~\cite{larios}, in which $x_t$
should be $a_t$.}
\begin{eqnarray}
f_0 &=& \frac{a^2_t (1+x_0)}{a^2_t(1+x_0)+2(1+x_m+x_p)}\, ,
 \qquad
f_- = \frac{2(1+x_m)}{a^2_t(1+x_0)+2(1+x_m+x_p)}\, ,
\label{eqfp} \\
\Delta \sigma_t &=& a_0 x_0 +a_m x_m + a_p x_p + a_5 x_5\, ,
\qquad
 \Delta \sigma_s = b_0 x_0 +b_m x_m + b_p x_p + b_5 x_5\, ,
\label{sigmas}
\end{eqnarray}
where $\Delta \sigma$ stands for the variation from the SM NLO
prediction, and the degree of right-handed polarization of the $W$
boson from top decay is obtained from $f_+=1-f_--f_0$. The
numerical values of the $a_i$ and $b_i$ coefficients were given in
Ref.~\cite{larios} for the Tevatron and the LHC, respectively.
They were obtained by integrating over the parton luminosities
which are evaluated using CTEQ6L1 parton distribution functions.
Furthermore, in the above equations,
\begin{eqnarray}
x_0\; &=& (f_1^L + f_2^R/a_t)^2 + (f_1^R + f_2^L/a_t)^2 - 1\, ,
\qquad
 x_m\; =(f_1^L + a_t f_2^R)^2  - 1\, ,
\nonumber \\
x_p\; &=& (f_1^R + a_t f_2^L)^2\, ,
\qquad
 x_5\; = a_t^2 ({f_2^L}^2+{f_2^R}^2) \, ,
\qquad {\rm with} \quad a_t = m_t/M_W \,. \nonumber
\end{eqnarray}

In case that a new light resonance, either a scalar or vector
boson, is found, the s-channel process could be significantly
enhanced and its production rate may not be dominated by a virtual
$W$-boson (s-channel) diagram. The above formulas may also apply
to models with extra heavy fermion ($T$), such as the top quark
partner in the Little Higgs (LH) Models that couples to the SM $b$
quark and $W$ boson. The $TbW$ coupling in general has the same
form of the general $tbW$ coupling given above, and the
expressions for single-$T$ production cross sections are exactly
the same as single-top except for the heavy mass $m_{T}$. The size
of the coefficients in the production cross sections decrease
drastically with a greater mass $m_{T}$. For instance, at
$m_{T}\,=500$ GeV the $a_0$ coefficient decreases one order of
magnitude with respect to the value for $m_{T}\,=178$ GeV at the
LHC. Furthermore, in the t-channel single-$T$ process, the $a_0$
coefficient, corresponding to longitudinal $W$ boson contribution,
dominates its production cross section. This is in analogy with
the SM t-channel single-top production. in which longitudinal $W$
boson contribution dominates the inclusive cross section.

Finally, we note that with enough precision in determining the
four individual $tbW$ couplings at the LHC, we could start
distinguishing models of EWSB \cite{larios}, such as the MSSM and
the Technicolor assisted Topcolor (TC2) mode.

\section{ Top quark and EWSB}

\subsection{MSSM and TC2}

With its mass around the weak scale, top quark could very well be
a special quark that might play an important role in the EWSB. The
bottom-up approach is to construct the most general electroweak
chiral Lagrangian to describe the interaction of top quark and
then compare the effective theory predictions with experimental
data to extract out information on the coefficients of the
relevant effective operators. Since these coefficients depend on
the underlying new physics models, it is possible to discriminate
models of EWSB from their values. Usually, two classes of EWSB
models are considered in the literature: weakly interacting models
and strongly interacting models. The former consists of elementary
Higgs boson(s) that originate from spontaneously symmetry breaking
(such as MSSM and LH models), and the latter may predict composite
Higgs boson(s) due to some strongly interacting dynamical symmetry
breaking mechanism (such as TC2 models). In the MSSM, the EWSB is
generated radiatively due to heavy top quark contribution in
loops. In the TC2 model, top quark condensate relates the heavy
top quark mass with the weak scale ($v\sim 246$\,GeV) at which the
$W$ and $Z$ bosons become massive gauge bosons. It is interesting
to note that numerically $m_t \sim v/\sqrt{2} \sim M_W+M_Z$.

\begin{figure}
  \centering
     \begin{minipage}[c]{0.65\textwidth}
    \centering
    \includegraphics[scale=0.5]{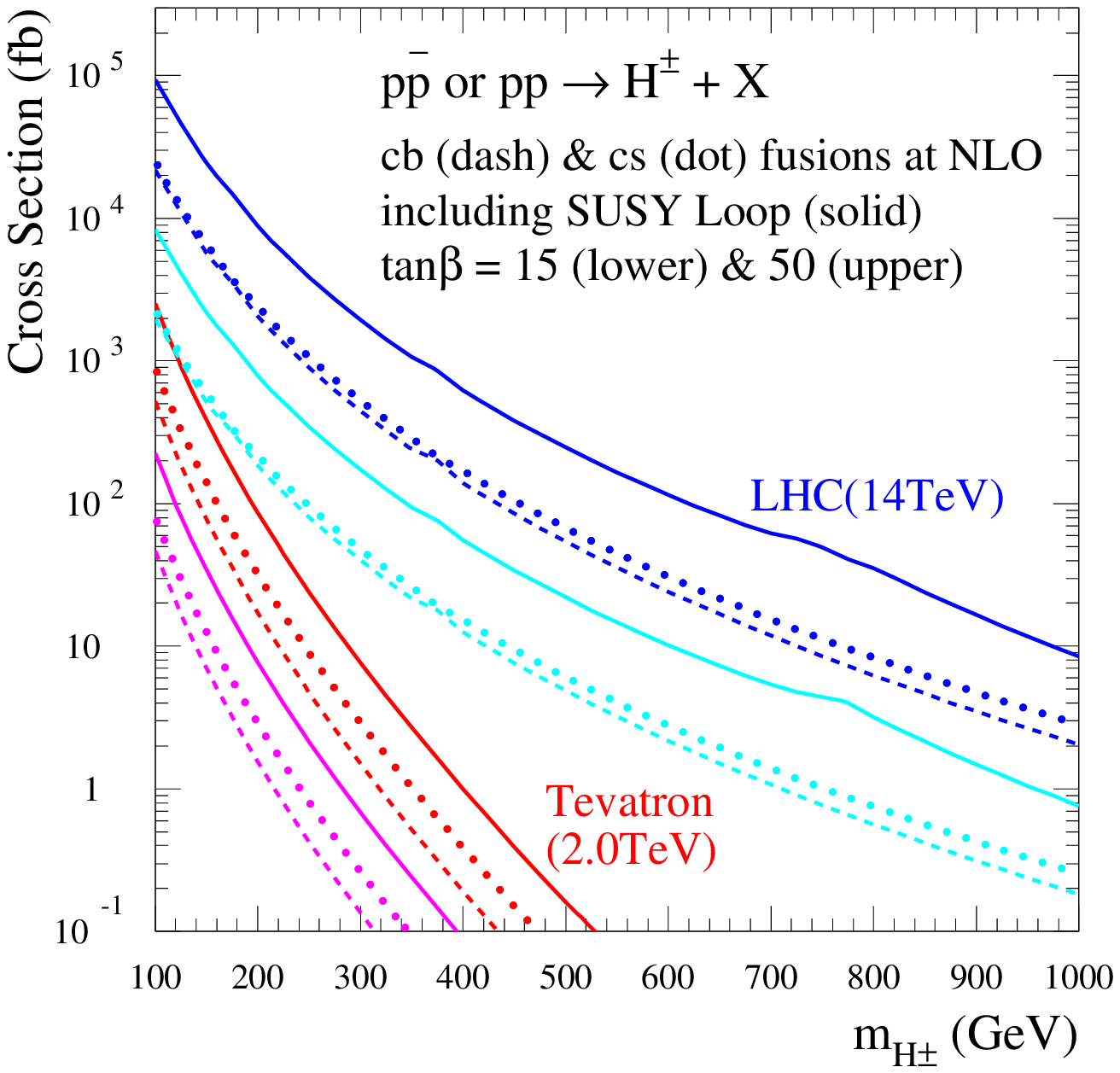}
    \end{minipage}%
    \begin{minipage}[c]{0.35\textwidth}
    \centering
    \includegraphics[scale=0.5]{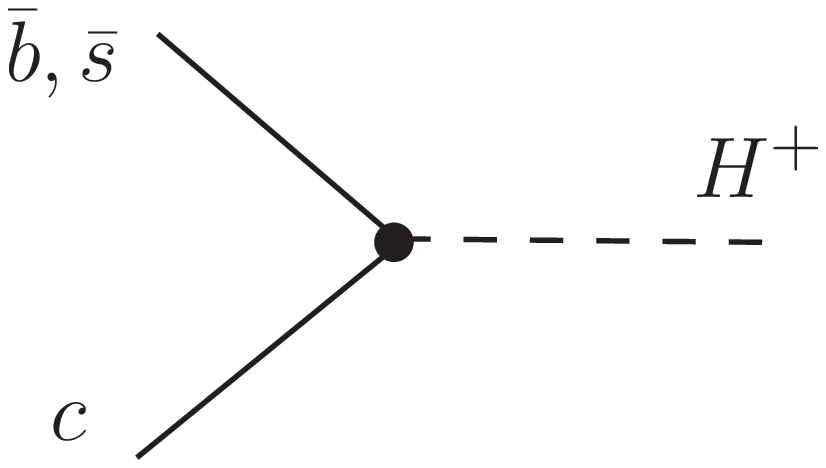}
    \end{minipage}
\caption{The MSSM $H^\pm$ production via $cb$ (and $cs$) fusions
at the Tevatron and the LHC.} \label{fig:bchp}
\end{figure}

Because bottom quark is the isospin partner of top quark, its
interactions can also be sensitive to new physics models of EWSB.
For example, in the MSSM, two Higgs doublets are required by
supersymmetry. When $\tan \beta$ (the ratio of vacuum expectation
values of the two Higgs doublets) is large, the bottom quark
Yukawa coupling can become large enough that the production rates
of $b{\bar b}H$ events can become measurable at the Tevatron and
the LHC \cite{bbh}. Assuming a horizontal U(1) flavor symmetry in
the soft-breaking sector of MSSM for generating the scalar quark
masses, the (32) and (23) entries of the trilinear coupling matrix
$A_u$ for up-squarks can have the same size as the (33) entry of
$A_u$ and lead to large flavor mixing between stop ($\tilde{t}$)
and scharm  ($\tilde{c}$) \cite{hjhe}. As shown in
Ref.~\cite{hjhe}, large $\tilde{t}$-$\tilde{c}$ mixing can enhance
the s-channel charged Higgs boson production via $c {\bar b} \to
H^+$, cf. Fig.~\ref{fig:bchp}, as well as the FCNC decay process
$h^0 \rightarrow t {\bar c}$. We note that the s-channel $H^+$ production
can be tested in the single-top events when $H^+$ decays into a
pair of $b$ and $W$ at the Tevatron and the LHC. The decay
branching ratio BR($h^0 \rightarrow t {\bar c}$) can range from $10^{-5}$
to $10^{-3}$ and is sensitive to the mass of the lightest stop and
the mixing of stop and scharm. Furthermore, the chirality of
$H^+$-$b$-$c$ couplings can be determined at future photon-photon
linear collider via $\gamma \gamma \rightarrow (b {\bar b})(c {\bar c})
\rightarrow H^+ b {\bar c}$ by polarizing the polarization of the incoming
photon beams \cite{shinya}.

Similarly, in the TC2 model, top-pion ($\pi^{\prime}$) can be
produced at resonance via the s-channel process $c {\bar b} \to
\pi^{\prime} \rightarrow t {\bar b}$ due to the presence of a large
$t_R-c_R$ mixing that is consistent with all the existing FCNC
data \cite{hjhetc}. We note that the CKM matrix is determined by
the product of the left-handed up- and down-type quark rotation
matrices (to diagonalize the quark mass matrices) and the
right-handed rotation matrices are not fixed by the CKM matrix.
Hence, a large $t_R$-$c_R$ mixing does not contradict with the
existing experimental data. In the Topflavor model \cite{ehab},
the production of an s-channel $W'$ heavy boson can also produce
single-top signature through $q {\bar q^{\prime}} \rightarrow W' \rightarrow t
{\bar b}$ at hadron colliders.

In summary, carefully studying the interactions of top quark in
experimental data can help distinguishing models of EWSB. A few
examples are given in Table~\ref{tab:ewsb} to illustrate the idea.

\begin{table}
\centering
\begin{tabular}{c|c|c}
\hline {model} & {top Yukawa coupling} & {bottom Yukawa coupling}
\tabularnewline \hline SM & $\sim 1$ &  $\sim 1/40$
\tabularnewline \hline MSSM ($\tan \beta =40$)  & $\sim 1/40$ &
$\sim 1$ \tabularnewline \hline TC2 & $\sim 1$ &  $\sim 1$
\tabularnewline \hline
\end{tabular}

\caption{Discriminate models of EWSB by testing the interaction of
top and bottom quarks to Higgs boson. \label{tab:ewsb}}
\end{table}

\subsection{Little Higgs model}

In the Little Higgs model, the Higgs boson mass is naturally at
the weak scale, because large quadratic correction to the Higgs
boson mass term induced by the top quark loop is cancelled by the
fermionic partner ($T$) of top quark ($t$) due to the approximate
global symmetry which relates T with t (i.e., Little Higgs
mechanism)~\cite{lht}. Furthermore, to ensure $\rho$-parameter to
be one at tree level, a discrete symmetry called T-parity was
introduced in the Little Higgs model with T-parity (LHT).
Consequently, the effective cutoff scale of the model $\Lambda=4
\pi f$ can be as low as 10\,TeV and the masses of new heavy
resonances can be of sub-TeV \cite{lht-t}. The LHT model is
particularly interesting because it also provides a dark matter
candidate which is the lightest T-odd particle $A_H$, the heavy
bosonic T-partner of photon. Another important feature of this
model is that new Higgs couplings are induced in the part of
effective theory that generates the masses of the extra heavy
T-partner (either T-odd or T-even) fermions needed for protecting
the Higgs boson mass at the weak scale. In Ref.~\cite{tobe} we
showed that these new Higgs couplings can lead to non-decoupling
effect and alter our conclusions on the collider phenomenology of
Higgs boson.

\begin{figure}
\begin{center}
\includegraphics[%
  scale=0.4]{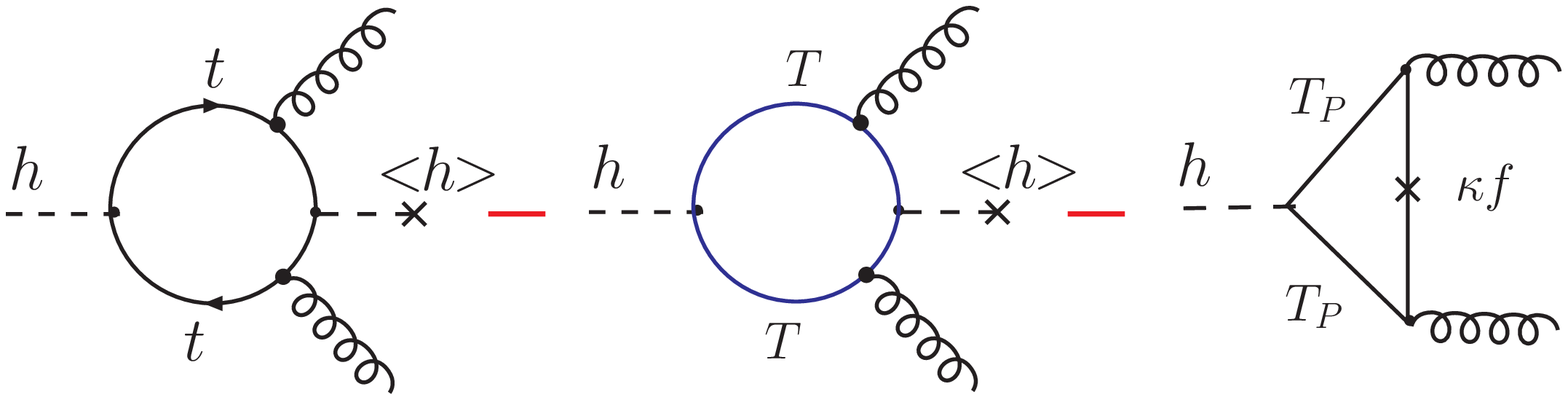}
\end{center}

\caption{Representative Feynman diagrams for $gg \rightarrow h$
production.} \label{fig:ggh}
\end{figure}

For example, the tree level couplings of Higgs boson to weak gauge
bosons and fermions are all suppressed relative to their SM values
by a factor $1- c (v_{SM}^2/f^2)$ where $v_{SM} \sim 246$\,GeV and
the coefficient $c$ depends on the specific coupling and model
scenario. In Fig.~\ref{fig:ggh}, we show some representative
Feynman diagrams contributing to the production process $gg \to
h$. We found that the production rate of Higgs boson via
gluon-gluon fusion is also suppressed relative to the SM rate.
This can be understood as follows. In the Littlest Higgs model
\cite{littlest}, the contribution from the T-partner ($T$) of top
quark partially cancel the top quark loop contribution, similar to
the effect of cancelling the quadratic divergencies in Higgs boson
mass correction. The additional contribution induced by the T-odd
heavy fermions further suppress the production rate of $gg \rightarrow h$
due to the non-decoupling effect originated from the mass
generation mechanism for those heavy T-odd fermions ~\cite{tobe}.
With $f=700$\,GeV, which is consistent with low energy precision
data, the cross section $\sigma(gg \rightarrow h)$ can be reduced by about
35\% for $m_h$ around 115\,GeV. It is important to note that the
total decay width of Higgs boson in the LHT is always smaller than
that predicted by the SM, for the cancellation of quadratic
divergencies in Higgs boson mass corrections are among particles
with the same spin statistics. While the partial decay width of
$\Gamma(h \rightarrow \gamma \gamma)$ does not change very much from the
SM prediction, the decay branching ratio BR($h \rightarrow \gamma \gamma
$) can increase by as much as 30\% for the total decay width of
Higgs boson is largely reduced by the smaller bottom quark Yukawa
coupling. Consequently, the discovery potential of the LHC for a
light Higgs boson with mass around 100\,GeV can be dramatically
altered from its usual conclusion. For instance, the rate of $gg
\rightarrow h \rightarrow \gamma\gamma$ can be reduced by about 14\%, while the
$W$-boson fusion rate increases by about 22\% as compared to the
SM prediction, cf. Table 1 of Ref.~\cite{tobe}. Hence, in the LHT
model, the $W$-boson fusion process becomes the main discovery
channel of a 100\,GeV Higgs boson at the LHC \cite{tobe}.

\subsection{Higgsless model}

\begin{figure}
\begin{center}
\includegraphics[%
  scale=0.5]{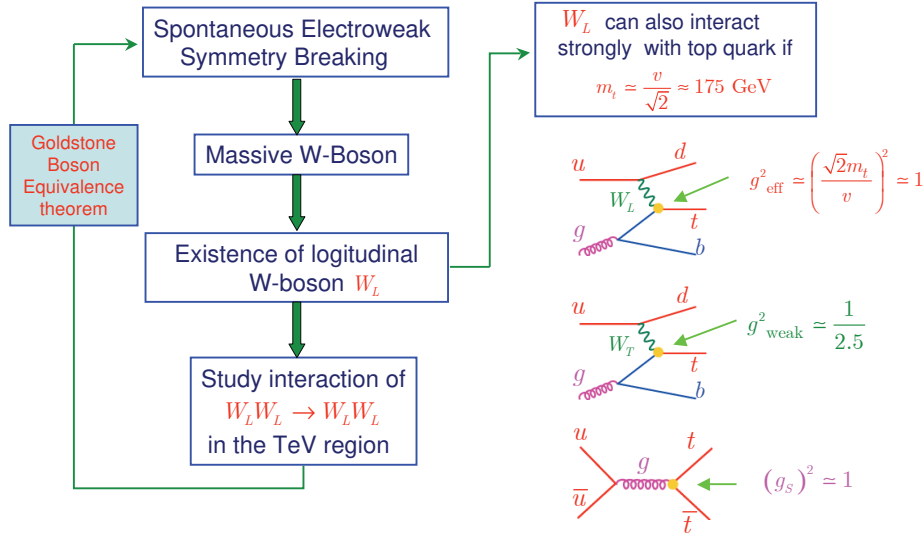}
\end{center}

\caption{Schematical diagrams to show the motivation of study done
in Ref.~\cite{st1990} for t-channel single-top production. }
\label{fig:wwst}
\end{figure}

If Higgs boson exists, discovering the Higgs boson and studying
its interaction is essential to probe the electroweak symmetry
breaking and the flavor symmetry breaking. Otherwise, we need to
carefully study the interaction among longitudinal $W$
(representing $W^\pm$ or $Z$) bosons ($W_L$) in the TeV region
\cite{lhcwwww} as well as the interaction of longitudinal $W$
boson to heavy fermions (top and bottom) \cite{larioswwtt}. What
motivated my 1990 single-top paper was to find a new way to study
the interaction of longitudinal $W$ boson to heavy top quark (with
its mass around 180\,GeV) \cite{st1990}. Schematically, the
relation between the $W_LW_L \rightarrow W_LW_L$ scattering and the
t-channel single-top process is shown in Fig.~\ref{fig:wwst}.

In the effective-$W$ approximation, the t-channel single-top
production is dominated by the interaction of longitudinal $W$
boson ($W_L$) to heavy top. While the interaction of transverse
$W$ boson ($W_T$) to top quark has the typical weak coupling
strength, the interaction of $W_L$ and $t$ is effectively of the
same strength as $O(1)$ Yukawa coupling of top quark based on the
Goldstone Equivalence Theorem \cite{et}.

In some models of extra-dimension, say, Higgsless model, there is
neither elementary nor composite Higgs boson to regulate the bad
high energy behavior of the $W_LW_L \rightarrow W_LW_L$ scattering
amplitudes in the TeV region. The breakdown of the unitarity of
scattering amplitudes is delayed by the extra heavy Kaluza-Klein
gauge bosons predicted in the four-dimensional effective theory
\cite{higgsless}. In such kind of models, the interaction of top
quark (and its partners) with longitudinal $W$ bosons can become
more important. Hence, we need to study the production of top
quark (and its partners) from $W$-boson fusion process in the TeV
region. It is also possible that the extra bosons needed for
delaying the unitarity breakdown in the $W_LW_L \rightarrow W_LW_L$
scattering amplitudes can lead to interesting top quark
phenomenology. For example, $Z'$ can modify the $t \bar t$ event
distributions and $W'$ can induce extra production rate for
single-top events at the LHC.

\section{Summary}

Because of its heavy mass, top quark may very well be a special
quark that could provide hints on electroweak (and flavor)
symmetry breaking mechanism. Though theorists have explored many
probable ideas in which top quark plays a crucial role in the
construction of the theory model, it is entirely up to the
experimental data to tell us which theory model is the closest to
true Nature. Needless to say that we need experimental data to
advance our acknowledge. Currently, we are waiting for the
exciting new discovery of single-top events at the Tevatron
Run-II. With the huge production rate of top quarks at the LHC, we
are able to study many details about its property. The physics of
top quark is indeed very rich, and it is also a sure thing that we
will learn much more with the realization of future ILC.

\acknowledgments

I would like to thank the local organizers for their hospitality
and for making the Workshop to be a productive one. I also thank
my collaborators in the past years who helped me to discover the
beauty in physics of top quark and to invent new ways to study its
phenomenology. They are G.L. Kane, G.A. Ladinsky, D.O. Carlson, E.
Malkawi, T.M.P. Tait, F. Larios, H.-J. He, L. Diaz-Cruz, S.
Mrenna, Q.-H. Cao, K. Tobe, and C.-R. Chen. I apologize that due
to the limited space in this write-up, I am not being able to cite
all the references in the literature that are relevant to top
quark physics, but they can be found collectively in the upcoming
TeV4LHC report. This work is supported in part by the U. S.
National Science Foundation under award PHY-0244919.

\end{document}